\documentclass[twocolumn,letter]{jpsj2}
\usepackage{graphicx}

\title{Phase Diagram of Spinless Fermions on an Anisotropic Triangular Lattice at Half-filling}

\author{
Chisa Hotta
\thanks{E-mail address:chisa@phys.aoyama.ac.jp}, Nobuo Furukawa, Akihiko Nakagawa, Kenn Kubo}
\inst{%
Aoyama Gakuin University, Fuchinobe 5-10-1, Sagamihara-shi,  
Kanagawa 229-8558, Japan
}
\recdate{\today}
\abst{
The strong coupling phase diagram of the spinless fermions on the anisotropic 
triangular lattice at half-filling is presented. 
The geometry of inter-site Coulomb interactions rules the phase diagram. 
Unconventional charge ordered phases are detected 
which are the recently reported pinball liquid and the striped chains. 
Both are induced by the quantum dynamics out of classical disordered states 
and afford extremely correlated metallic states and 
the particular domain wall-type of excitations, respectively. 
The disorder once killed by the quantum effect 
revives at the finite temperature, 
which is discussed in the terms of the organic $\theta$-ET$_2X$. 
}
\kword{%
}
\begin{document}
\maketitle
Charge degrees of freedom provide 
particular orders and disorders under geometrical frustration, 
as has been discussed from the time of the Verwey transitions\cite{verway}. 
Many recent topics are related to those on the anisotropic 
triangular lattice(ATL). 
They are experimentally explored in the organic solids as 
a possible "non-magnetic" Mott insulator\cite{shimizu} and 
organic thyristor\cite{terasaki}. 
Theoretically, 
the phase diagram of the half-filled Hubbard model on ATL 
seems to suggest the particular ``spin liquid'' insulator\cite{imada}, 
which has been searched for from the past 
as a nonmagnetic state free from any kind of ordering\cite{anderson}. 
\par
Quarter-filling is characterized by the charge ordering(CO) phenomenon, 
as studied by the extended Hubbard model(EHM) 
in the context of organic solids\cite{chemrev,chisa}. 
The interplay of CO and frustration is investigated 
in the phase diagrams based on ATL\cite{merino,hwata}. 
However, the two dimensional systems under geometrical frustration 
requires a delicate and difficult numerical treatment, 
thus a clear answer is still not obtained. 
In the present letter, we simplify the problem and 
deal with the spinless fermions at half-filling 
to extract the intrinsic character of the charge degrees of freedom. 
We obtain a convincing phase diagram which retains a strong coupling 
nature even in a rather weak coupling region. 
In addition to the pinball liquid reported in Ref.[10], 
another unconventional strongly correlated CO state with fluctuating stripes 
is found in the finite size cluster. 
Its transport character is sensitive to the geometry of 
quantum dynamics which is discussed in terms of the organic solid $\theta$-ET$_2X$. 
\par
\begin{figure}[t]
\begin{center}
\includegraphics[width=7cm]{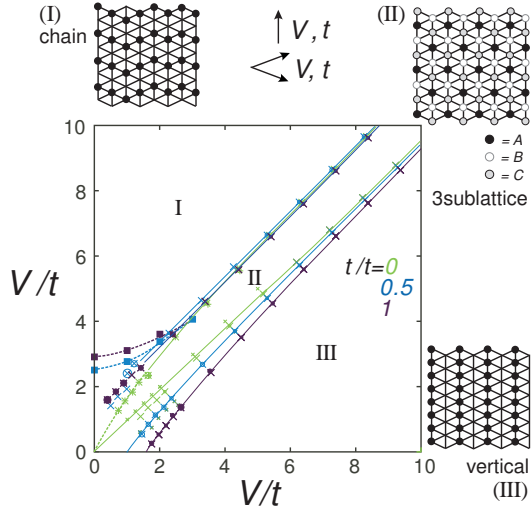}
\end{center}
\caption{
Ground state phase diagram of eq.(\ref{tvham}) on the plane of $V/t$ and $V'/t$ 
obtained numerically at $N$=24. 
It is classified into three regions, (I) chain stripe, (II) pinball liquid and 
(III) vertical stripe. 
The phase boundary is given at $t'/t$=1, 0.5 and 0. 
Broken lines with square symbols indicate the upper bound of the insulating state 
in (I) for $t/t'=0.5,1$. 
Shaded region includes the possible location of $\theta$-ET$_2X$. 
Representative configuration of the classical limit of the three phases are schematically 
shown together. 
The arrows with $V,t$ and $V',t'$ at the top indicate the directions of 
interacting bonds in eq.(1). 
}
\label{f1}
\end{figure}

The spinless fermion Hamiltonian in the present study is given as, 
\begin{equation}
\begin{split}
{\cal H} = \sum_{\langle i,j \rangle} &\Big(
         -t_{ij}c^\dagger_i c_j + {\rm h.c.} + V_{ij} n_i n_j\Big). 
\label{tvham}
\end{split}
\end{equation}
Here, $c_j$ and $n_j$ denote the annihilation and number operator of fermions. 
The triangular lattice has anisotropy in one of three directions as reflected 
in the transfer integrals, $t_{ij}$=$t,t'$, and the inter-site Coulomb interactions, 
$V_{ij}=V,V'$, between nearest neighbor sites 
whose corresponding bond directions are defined in Fig.\ref{f1}. 
The phenomena realized in EHM in the temperature range of 
$J\!\sim \!\frac{t^2}{U}\! < T\! \ll \!V$ 
are suitably described by the $t$-$V$ model. 
\par 
We perform the exact diagonalization on $N=4 \times 6=24$ cluster 
in eq.(\ref{tvham}). 
Figure~\ref{f1} shows the phase diagram on the plane of $V/t$ and $V'/t$ 
at several fixed values of $t'/t$ with periodic boundaries. 
The diagram is classified into three different regions, 
whose boundaries are estimated from 
the discontinuity in charge structural factor, 
the gradient of energy by $V$ and $V'$, 
and Drude-like component as we see shortly. 
The phases (I) and (III) are the two-fold periodic striped CO states 
whereas the phase (II) has a three fold periodicity 
and is denoted as "a pinball liquid" which is studied in Ref.[10]. 
The $t'/t$-dependence of the phase boundaries is very small. 
This phase diagram describes well 
the strong coupling physics at around $V,V'>2t$ 
below which the phase transition begins to smear.
\par
\begin{figure}[t]
\begin{center}
\includegraphics[width=5.5cm]{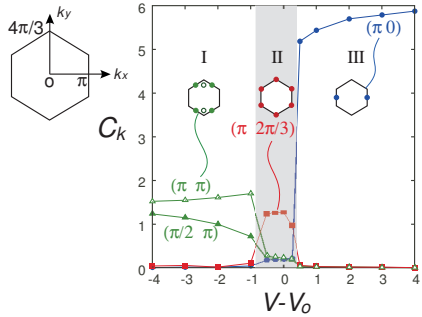}
\end{center}
\caption{Amplitude of charge structural factors along $V\!+\!V'\!=\!2V_o=\!12$ 
with characteristic wave numbers. 
Rectangle represents the Brillouin zone and 
the circles are the location of wave numbers with peaks. 
}
\label{f2}
\end{figure}

We first present the structural factor, 
$C_k=\frac{1}{N}\sum_{lm} \langle (n_l\!-\!1/2)(n_m\!-\!1/2)\rangle 
{\rm e}^{{\rm i}(r_l-r_m)k}$, in Fig.~\ref{f2}. 
Each phase is characterized by its 
own wave numbers,  
(I)$(\frac{\pi}{2}n,\pi)$ with integer $n$, 
(II)$(\pi,\pm\frac{2}{3}\pi),(0,\pm\frac{4}{3}\pi)$, and (III)$(\pi,0)$. 
The peak amplitudes vary discontinuously from one to another 
at the phase transitions, 
which confirm their first order character. 
We mention that the structural factors over 
the entire Brillouin zone beside each characteristic ones
are almost completely suppressed, 
which means that there is little possibility of the coexistence phase. 
The amplitudes in (I) and (II) are significantly 
smaller than the one in (III). 
As we see shortly, the disorder exists 
in the classical limit of phases (I) and (II), 
which still influences the states at $t \ne 0$ 
and suppresses the two-body correlation. 
\par
%%*%*%*%*%*%*%*%*%*%*%*%*%*%*%*%*%*%*%*%*%*%*%*%*%*%*%*%*%*%*%*%*
%%*%*%*%*%*%*%*%*%*%*%*%*%*%*%*%*%*%*%*%*%*%*%*%*%*%*%*%*%*%*%*%*
%*%*%*%*%*%*%*%*%*%*%*%*
%*%*%*%* fig 1 *%*%*%*%*
%*%*%*%*%*%*%*%*%*%*%*%*
%*%*%*%*%*%*%*%*%*%*%*%*
%*%*%*%*%*%*%*%*%*%*%*%*
%*%*%*%* fig 2 *%*%*%*%*
%*%*%*%*%*%*%*%*%*%*%*%*
%*%*%*%*%*%*%*%*%*%*%*%*
%*%*%*%*%*%*%*%*%*%*%*%*
%*%*%*%* fig 3 *%*%*%*%*
%*%*%*%*%*%*%*%*%*%*%*%*
\begin{figure}[t]
\begin{center}
\includegraphics[width=8cm]{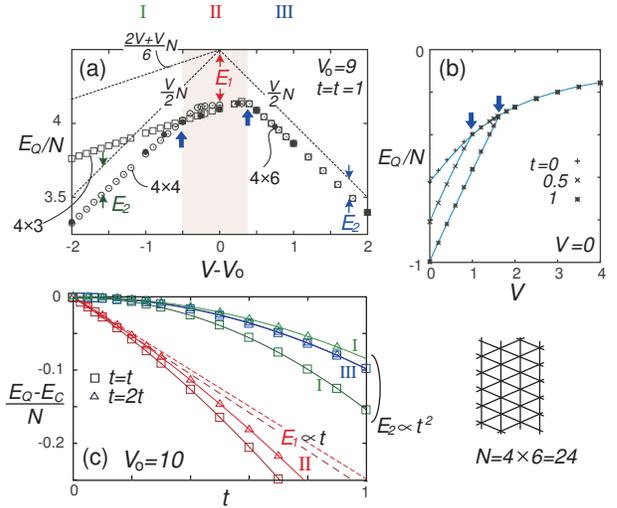}
\end{center}
\caption{Ground state energy of eq.(\ref{tvham}), $E_Q$, 
(a) at fixed $V+V'=2V_o=18$ with $t\!=\!t'=\!1$ 
under several cluster size. 
Broken lines give the classical binding energy per site, 
where $E_1,E_2$ are the energy gain per site 
due to $t,t'$. 
The $4 \times 6$ cluster shown together on the right. 
(b) $V'$-dependence of $E_Q$ at $V$=0. 
for several choices of $t'=0,0.5,1$ ($t=1$). 
Arrows indicate the phase transitions between (I) and (II). 
(c) Energy correction, $(E_Q-E_C)/N$, due to $t=t'$ and $t=2t'$ 
at $V_o$=20 in three phases, $V$=8(I), 10(II), and 12(III), 
which are assigned $E_2$, $E_1$, and $E_2$, respectively. 
}
\label{f3}
\end{figure}
%*%*%*%*%*%*%*%*%*%*%*%*
%*%*%*%*%*%*%*%*%*%*%*%*
%*%*%*%* fig 4 *%*%*%*%*
%*%*%*%*%*%*%*%*%*%*%*%*
\begin{figure}[t]
\begin{center}
\includegraphics[width=7cm]{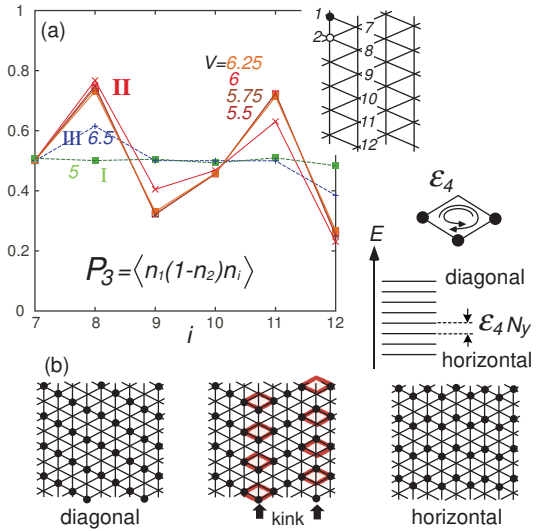}
\end{center}
\caption{(a) Three-body charge correlation function, $P_3(j)$, along the 
second chain($i=7-12$) at $t=t'=1$ for several choices of $V$ with $V_o=6$, 
with $i$ denotes the site index. 
Those in broken lines are the cases of striped states. 
(b) Fourth order process that lifts the degeneracy of the chain state. 
The cyclic permutation of three charges on the rectangle gives $\epsilon_4$. 
The energy spectrum becomes equally spaced by $\epsilon_4N_y$, 
with bottom and top as the horizontal and diagonal CO, respectively. 
The number kinks including $N_y$-rectangles in bold lines 
determines the energy. 
}
\label{f4}
\end{figure}
%*%*%*%*%*%*%*%*%*%*%*%*
%*%*%*%*%*%*%*%*%*%*%*%*
%*%*%*%*%*%*%*%*%*%*%*%*
%*%*%*%*%*%*%*%*%*%*%*%*
%%*%*%*%*%*%*%*%*%*%*%*%*%*%*%*%*%*%*%*%*%*%*%*%*%*%*%*%*%*%*%*%*
%%*%*%*%*%*%*%*%*%*%*%*%*%*%*%*%*%*%*%*%*%*%*%*%*%*%*%*%*%*%*%*%*
%%*%*%*%*%*%*%*%*%*%*%*%*%*%*%*%*%*%*%*%*%*%*%*%*%*%*%*%*%*%*%*%*
Figure\ref{f3}(a) shows the ground state energy, $E_{Q}$, which behaves 
almost linearly with $V$ at the fixed value of $V+V'\equiv 2V_0$, 
while its gradient changes discontinuously at the phase transition. 
It is also seen in Fig.\ref{f3}(b). 
Therefore the phase transitions are of first order. 
Note that the present phase diagram is almost free from finite size effect. 
Figure \ref{f3}(a) reveals that $E_Q$ of the smaller cluster size, $4\times 4$ and 
$4\times 3$, reproduces within 0.3$\%$ 
the ones of $4\times 6$ in (I), (III) and (II), (III), respectively, 
i.e. as long as the periodicity of each state is 
commensurate with the shape of the cluster. 
\par
The difference between three phases can be clearly understood 
by the strong coupling theory\cite{pinball}. 
In the classical limit $t=t'=0$, the ratio of $V$ and $V'$ determines 
the phase diagram. 
The system has a semimacroscopically ($\sim 2^{\sqrt{N}}$-fold) degenerate chain CO state 
and a non-degenerate vertical stripe CO state 
at $V<V'$ and $V>V'$, respectively\cite{pinball}, 
and the phases (I) and (III) are their extensions. 
Their lowest order energy gain per site by $t,t'$ ($\ll V,V'$) is estimated as, 
$E_2=-t'^2/V'-t^2/(2V'-V)$ (I), $-2t^2/(3V-2V')$ (II), 
which is presented numerically in Fig.~\ref{f3}(a) as $(E_Q-E_C)/N$. 
Since this effect is small at $V,V'\gg t,t'$, the stripes are quite robust 
even though the charge localization length is somewhat extended. 
\par
When $V=V'$ the three sublattice state with a macroscopic degeneracy joins 
the ground state besides these two stripes in the classical limit. 
The quantum dynamics of $t$ by first order lifts the degeneracy of 
part of the three sublattice states and the pinball liquid (II) appears. 
The estimated kinetic energy gain ($(E_Q-E_C)/N$ in Fig.~\ref{f3}(a)) 
is of first order ($t$-linear) 
which behaves as $E_1 \propto c_1t + c_2t^2$ ($c_1,c_2$ are constants) 
in contrast to $E_2 \propto t^2$ as shown in Fig.~\ref{f3}(c) 
regardless of the anisotropy in $t'/t$. 
Since $E_Q$= $V/2\!+\!E_2$, $(2V'+V)/6\!+\!E_1$, and $V/2\!+\!E_2$ for (I)-(III), 
respectively, 
the phase (II) appears at around $6(E_1-E_2) \leq V-V'\leq -3(E_1-E_2)$. 
\par
To fully figure out the nature of the pinball liquid (II) out of disorder, 
we should consider the three-body correlation, 
$P_3(j)\!\equiv \!\langle n_1 (1-n_2) n_j \rangle$ 
which is the population of the $j$-th site 
when the {\it 1}-st and the {\it 2}-nd site is present and absent, 
respectively. 
Figure~\ref{f4}(a) shows that 
the $n_A+n_B\! \sim \!2n_C$ type of three-fold structure is 
present only in (II). 
This is in contrast to the $n_A \neq n_B\sim n_C$ type ones 
based on mean field wave functions\cite{hwata}. 
\par
The phase (I) is another ordered state out of disorder. 
Its classical limit has a CO along every chain ($y$-direction). 
Regular stacking of the CO chains give 
the diagonal striped structure shown in Fig.~\ref{f4}(b). 
Starting from this we can introduce a kink one by one, 
and finally we obtain the horizontal stripe with 
the full number of kinks. 
All these structures have the same energy and are 
semi-macroscopically degenerate. 
We call this gapless and disordered state a chain CO\cite{pinball}. 
The energy correction by $t$ starts from the second order, but 
the degeneracy is not lifted until the fourth order; 
the degree of lifting per rectangular unit with three charges is 
$\epsilon_4 \!= \!-2t^4/(2V'-V)^2V'$ as shown in Fig.~\ref{f4}(b). 
The energy gain from the diagonal stripe is given as 
$\epsilon_4 N_y N_{\rm kink}$, where $N_{\rm kink}$ 
denotes the number of kinks. 
Therefore the horizontal stripe 
becomes a unique ground state in the bulk limit, 
and the reduction of $N_{\rm kink}$ 
corresponds to the gapful domain wall excitation 
which requires energy larger than the 
single particle excitation($\epsilon_4 N_y \!\gg \!V'$) 
except in the vicinity of the strong coupling limit. 
In finite clusters, however, these kinked chain states easily 
mix with the horizontal stripe. The energy gain here is, 
$\epsilon_{\rm ch} \sim N_y t'^{\frac{N_y}{2}}/V'^{\frac{N_y}{2}-1}$, 
which is far smaller than $E_1$ and $E_2$ at $V' \gg t'$ 
so that its correction to the phase boundary in Fig.~\ref{f1} is negligibly small. 
In this way, the degeneracy lifted ground state 
in the bulk limit should have only a single peak of $C_k$ 
within the Brillouin zone at $(0,\pi)$. 
\par
%*%*%*%*%*%*%*%*%*%*%*%*
%*%*%*%* fig 5 *%*%*%*%*
%*%*%*%*%*%*%*%*%*%*%*%*
\begin{figure}[t]
\begin{center}
\includegraphics[width=7.5cm]{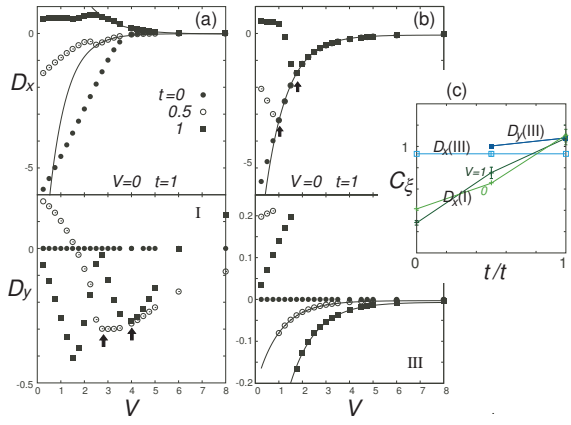}
\end{center}
\caption{
Amplitude of energy, $D_\alpha$, under the twisted boundary condition in 
$\alpha=x,y$, directions 
along (a)$V$=0 and (b)$V'$=0. 
Solid lines show the exponential fit, 
$D_\alpha \propto {\rm e}^{-\Delta_c/C_{\xi_\alpha}}$. 
Arrows indicate the phase transitions
(those in (a) are the possible metal-insulator transition points). 
The corresponding fitting constants, $C_{\xi_\alpha}$, are given in (c) 
as functions of $t'/t$.  
}
\label{f5}
\end{figure}
%*%*%*%*%*%*%*%*%*%*%*%*
%*%*%*%*%*%*%*%*%*%*%*%*
%*%*%*%* fig 6 *%*%*%*%*
%*%*%*%*%*%*%*%*%*%*%*%*
\begin{figure}[tbp]
\begin{center}
\includegraphics[width=7cm]{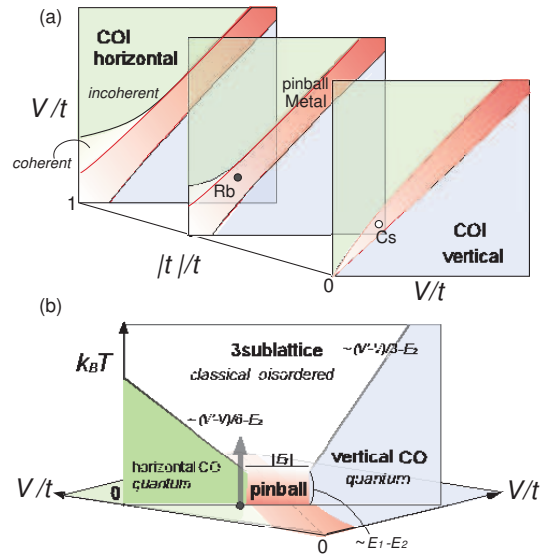}
\end{center}
\caption{(a) Schematic ground state phase diagram on the planes of 
$V/t$ and $V'/t$ at each $t'/t$. 
The COI denote the insulating striped phases. 
%The phase diagram is reliable at least 
%in the strong coupling region where the phase boundaries 
%follow the linear lines($V,V'>2t$ in Fig.~\ref{f1}). 
Filled and open circles are the possible locations of the 
Rb and Cs-salts of $\theta$-ET$_2X$. 
(b) Schematic phase diagram at finite temperature. 
The colored regions denote the quantum phases, (I)-(III), 
which are replaced by the classical disordered phase 
at higher temperatures. 
The crossover is possibly realized in the Cs-salt 
at $T_{\rm cr}$ 
as indicated in arrows. 
}
\label{f6}
\end{figure}
%*%*%*%*%*%*%*%*%*%*%*%*
%*%*%*%*%*%*%*%*%*%*%*%*
%*%*%*%*%*%*%*%*%*%*%*%*
%*%*%*%*%*%*%*%*%*%*%*%*%*%*%*%*%*%*%*%*%*%*%*%*%*%*%*%*%*%*
%*%*%*%*%*%*%*%*%*%*%*%*%*%*%*%*%*%*%*%*%*%*%*%*%*%*%*%*%*%*
%*%*%*%*%*%*%*%*%*%*%*%*%*%*%*%*%*%*%*%*%*%*%*%*%*%*%*%*%*%*
Next, we twist the boundary condition in $x$ and $y$ directions 
by the phase factor of e$^{{\rm i} \psi_\alpha}$ ($\alpha=x,y$), 
where $\psi_\alpha$=0 and $\pi$ correspond to the periodic and anti-periodic 
boundary conditions, respectively. 
Then, $D_\alpha\!=\!E_Q(\psi_\alpha\!=\!\pi)\!-\!E_Q(\psi_\alpha\!=\!0)$, 
is approximately proportional to the Drude weight of the finite system size. 
The discontinuity of $D_\alpha$ detects the 
first order phase transitions among (I) (II) and (III).
The pinball liquid is possibly a metal in the bulk limit 
with $D_\alpha \!\neq \!0$ as previously reported in Ref.[10]\cite{pinball}. 
In contrast, the coherence of the gapful localized commensurate stripes 
is expected to decay exponentially with distance $l$ as, 
$\propto {\rm e}^{-l/\xi}$, where $\xi$ denotes the coherence length. 
Then we except, $\xi_\alpha=C_{\xi_\alpha}/\Delta_c$ with 
$C_{\xi_\alpha}$ as $V,V'$-independent but $t',t$-dependent values 
and $\Delta_c$ the value of charge gap in the classical limit, 
i.e. $V'$ in (II) and $2V-V'$ in (III). 
By anticipating $D_\alpha \propto {\rm e}^{-N_\alpha C_{\xi_\alpha}/\Delta_c}$ 
($N_\alpha$ is the system length), 
the fitting is well performed as the solid lines in Figs.~\ref{f5}(a) 
and \ref{f5}(b) 
whose $C_{\xi_\alpha}$'s are plotted against $t'/t$ 
in Fig.~\ref{f5}(c). 
The fitting breaks down simultaneously at 
the transition from (III) to (II), 
which indicates that phase (III) is a localized ($\xi_\alpha < N_\alpha$) 
insulator. In this phase 
both $D_\alpha$ and $C_{\xi_\alpha}$ are $t'/t$-independent 
which is also the case with $E_Q$ in Fig.~\ref{f3}(b). 
In contrast, $E_Q$ of (I) in Fig.~\ref{f3}(d) 
together with $D_\alpha$ actually has 
large $t'/t$-dependence. 
There is a possible phase boundary within phase (I) 
estimated by the discontinuity in $\partial D_\alpha/\partial V'$ 
which is given in Fig.~\ref{f1} with broken lines. 
The strong coupling large $V'$-region is an insulator 
in its bulk limit. 
However, in the rather weak coupling region, 
$\xi_\alpha$ develops and exceeds $N_\alpha$, 
which corresponds to the above phase boundary. 
Notice that it depends much on $t'/t$. 
At present, we cannot discriminate whether this new intermediate 
region in (I) is gapful or not in the bulk case, nor 
whether there is a finite size effect in this boundary. 
Nevertheless, 
the estimated boundary gives an upper bound of the metal 
insulator transition line. 
\par
The previous theoretical attempts to clarify the phase diagram 
suffers from problems. 
Those based on mean-field wave function\cite{hwata} 
failed to describe fully the extremely correlated 
chain(horizontal) and pinball states. 
Their phase diagram has ambiguous "competition" region 
between states of different symmetry breaking due to large 
size-dependence, which is an artifact of adopting a 
plane wave based wave function on such correlated and 
highly localized system. 
Exact diagonalization on EHM in 4$\times$4 finite clusters 
overestimates their "frustration induced metal" as well as 
underestimates the pinball liquid\cite{merino}, as clarified explicitly 
in Figs.~\ref{f1} and \ref{f3}(a). 
Our phase diagram presents a clear view to these problems 
and describes the intrinsic nature of $U\!=\!\infty$ EHM 
at quarter-filling since the basic character of 
three phases are free from statistics\cite{pinball}. 
\par
The ground state phase diagram is summarized in Fig.~\ref{f6}(a). 
The anisotropy of $t'/t$ only affects the coherence of phase (I) 
in a relatively weakly coupled region. 
We show schematically in Fig.~\ref{f6}(b) 
the finite temperature effect, $k_BT\! \neq \!0$, 
where the classical disordered state 
becomes prosperous as a result of the macroscopic entropy. 
The free energy of the classical three-sublattice 
state is, 
$F_{\rm 3sub} \!\sim \!\frac {2V+V'}{6}N\!-\!k_BT \ln (0.5\times 2^N)$, 
while those of the nondegenerate quantum states are 
$F\!=\!E_Q \! \sim E_C+\delta E_Q N$ with 
$\delta E_Q \sim \!E_2+\epsilon_4$, $E_1$, and $E_2$ for 
(I), (II) and (III), respectively. 
The crossover lines from quantum to classical states 
with increasing $k_BT$ are estimated by the 
crossing points of these free energies as, 
$T_{\rm cr} \sim \frac{V'-V}{6}-E_2$, $|E_1|$ and 
$\frac{V-V'}{3}-E_2$, respectively. 
The chain type of disorder does not appear in the bulk case 
since the entropy gain remains 
of order $\sim O(\sqrt{N}\ln \sqrt{N})$.  
\par
Now we examine the relevance of the present results to 
the CO materials, $\theta$-ET$_2X$. 
The nature of this family at low temperatures is classified 
on the experimental phase diagram along 
the dihedral angle, $\phi$\cite{hatsumi};
when $\phi>110^\circ$, the system becomes a CO insulator 
accompanied by the structural transition at $T_{\rm CO}\sim 200$K. 
At $\phi <110^\circ $ it is a bad metal down to low temperatures 
with the resistivity minimum at $T_\rho \sim $ 20-70K. 
Recent X-ray analysis reports the existence of a short range diffuse spot of 
two-fold and three-fold type in the Cs-salt
($X$=Cs$M$(SCN)$_4$, $M$=Co,Zn, $\phi < 110^\circ$) 
at $\sim T_\rho$\cite{wat1} 
and also in Rb-salts($\phi \sim 110^\circ$) 
above $T_{\rm CO}$\cite{wat2}, 
which have $(|t'|/t,V'/V) \sim$ (0,0.9) and (0.5,0.85), respectively\cite{tmori}. 
\par
It is obvious that there is a characteristic crossover temperature 
of order $T_\rho$ at $\phi<110^\circ$ which increases with $\phi$. 
Actually, at around $V'>V$, we can estimate the 
crossover temperature as $T_{\rm cr} \sim \frac{V'-V}{6} \sim 0.1V$. 
Since we consider $V\!\sim \!T_{\rm CO}$, 
the experimental crossover at $\sim T_\rho$ is reasonably understood. 
An observed bad metallic nature with almost $T$-independent resistivity 
at $T>T_\rho$\cite{hatsumi} supports the existence of disorder.  
Also the up going of resistivity at $T<T_\rho$ is consistent 
with the insulating character of the possible horizontal stripe\cite{hatsumi}. 
The coexistence of states with different periodicity in the ground state 
is ruled out in the present study. 
However, at finite temperature, the disordered 
state which includes both the three-sublattice and 
the striped local patterns might appear around the crossover 
temperature, which were excluded in Fig.~\ref{f6}(b) for 
simplicity.
The study on this complicated matter is still underway. 
\par
Somehow, 
the ground state of this family is placed 
amidst the pinball liquid state as shown in Fig.~\ref{f1}, 
which is inconsistent with the above experimental findings. 
In the present stage we did not consider the effect of phonons. 
The three-fold state is found to be quite easily replaced 
by the horizontal stripe long range order(LRO) 
under that effect\cite{kaneko}, which might be valid in our case as well. 
\par 
It is frequently argued that $\phi$ can be regarded as $V/t$. 
Contrarily, one of the authors 
successfully reclassified the experimental phase diagram 
by $|t'|/t$, on the basis of EHM\cite{chisa} 
which is recently pursued experimentally\cite{kondo} and 
also in the weak coupling theory\cite{hwata}. 
In our phase diagram, however, 
$|t'|/t$ only rules the coherence 
length of the system at the relatively weak coupling region. 
To figure this out, 
we should examine the extra 
effect such as electron-phonon coupling 
in the disordered region. 
\par
In conclusion, the strong coupling phase diagram of the charges 
on the anisotropic triangular lattice is presented. 
A metallic pinball liquid phase 
appears in between an insulating vertical and 
the horizontal stripe CO's. 
The first order transitions take place in the phase boundaries which 
are almost solely determined by the geometry of the inter-site Coulomb interactions. 
The present phase diagram provides 
two different types of order-from-disorder due to quantum dynamics; 
the extremely correlated pinball liquid out of disordered 
three sublattice classical state, 
and the horizontal state which has a domain-wall like excitation 
to the disordered chain striped classical states. 
Both of these states provide good test cases 
for the disorder-embedded phenomena. 
The disorder once gave path to quantum states 
recovers at finite temperature with the aid of its macroscopic entropy. 
\par
We thank R. Kondo, M. Watanabe, I. Terasaki, and R. Chiba 
for many useful experimental informations. 
\vspace{-5mm}
%*%*%*%*%*%*%*%*%*%*%*%*%*%*%*%*%*
%*%*%*%*%*%*%*%*%*%*%*%*%*%*%*%*%*

\end{document}